\documentclass[superscriptaddress,twocolumn,showpacs,preprintnumbers,amsmath,amssymbnofootinbib]{revtex4}

\usepackage{graphicx}
\usepackage{dcolumn}
\usepackage{bm}

\begin{document}

\preprint{APS/123-QED}

\title{Prolate-Spherical Shape Coexistence at N=28 in $^{44}$S}

\author{C.~Force}\affiliation{Grand Acc\'el\'erateur National d'Ions Lours (GANIL), CEA/DSM-CNRS/IN2P3, Caen, France}
\author{S.~Gr\'evy}\email{grevy@in2p3.fr}\affiliation{Grand Acc\'el\'erateur National d'Ions Lours (GANIL), CEA/DSM-CNRS/IN2P3, Caen, France}
\author{L.~Gaudefroy}\affiliation{CEA,DAM,DIF, F-91297 Arpajon, France}
\author{O.~Sorlin}\author{L.~Caceres} \affiliation{Grand Acc\'el\'erateur National d'Ions Lours (GANIL), CEA/DSM-CNRS/IN2P3, Caen, France}\author{F.~Rotaru}\affiliation{Institute of Atomic Physics, IFIN-HH, Bucharest-Magurele, P.O. Box MG6, Romania}\author{J.~Mrazek}\affiliation{Nuclear Physics Institute, AS CR, CZ-25068 Rez, Czech Republic}\author{N.~L.~Achouri}\author{J.~C.~Ang\'elique}\affiliation{LPC Caen, ENSICAEN, Universit\'e de Caen, CNRS/IN2P3, Caen, France}\author{F.~Azaiez}\affiliation{IPNO, Universit\'e Paris-Sud 11, CNRS/IN2P3, Orsay, France} \author{B.~Bastin}\affiliation{LPC Caen, ENSICAEN, Universit\'e de Caen, CNRS/IN2P3, Caen, France}\author{R.~Borcea}\author{A. Buta}\affiliation{Institute of Atomic Physics, IFIN-HH, Bucharest-Magurele, P.O. Box MG6, Romania}\author{J. M. Daugas}\affiliation{CEA,DAM,DIF, F-91297 Arpajon, France}\author{Z.~Dlouhy}\affiliation{Nuclear Physics Institute, AS CR, CZ-25068 Rez, Czech Republic}\author{Zs. Dombr\'adi}\affiliation{Institute of nuclear Research, H-4001 Debrecen, Pf.51, Hungary}\author{F.~De~Oliveira}\affiliation{Grand Acc\'el\'erateur National d'Ions Lours (GANIL), CEA/DSM-CNRS/IN2P3, Caen, France}\author{F.~Negoita}\affiliation{Institute of Atomic Physics, IFIN-HH, Bucharest-Magurele, P.O. Box MG6, Romania}\author{Y.~Penionzhkevich}\affiliation{FLNR, JINR, 141980 Dubna, Moscow region, Russia}\author{M.~G.~Saint-Laurent}\affiliation{Grand Acc\'el\'erateur National d'Ions Lours (GANIL), CEA/DSM-CNRS/IN2P3, Caen, France}\author{D.~Sohler}\affiliation{Institute of nuclear Research, H-4001 Debrecen, Pf.51, Hungary}\author{M.~Stanoiu}\affiliation{Institute of Atomic Physics, IFIN-HH, Bucharest-Magurele, P.O. Box MG6, Romania}\author{I.~Stefan}\affiliation{Grand Acc\'el\'erateur National d'Ions Lours (GANIL), CEA/DSM-CNRS/IN2P3, Caen, France}\author{C.~Stodel}\affiliation{Grand Acc\'el\'erateur National d'Ions Lours (GANIL), CEA/DSM-CNRS/IN2P3, Caen, France}\author{F.~Nowacki}\affiliation{IPHC, CNRS/IN2P3 and Universit\'e de Strasbourg, F-67037 Stasbourg Cedex 2, France}

\date{\today}

\begin{abstract}
The structure of $^{44}$S has been studied using delayed $\gamma$ and
electron spectroscopy at \textsc{ganil}. The decay rates of the 0$^+_2$ isomeric state
to the 2$^+_1$ and 0$^+_1$ states
have been measured for the first time, leading to a reduced
transition probability B(E2~:~2$^{+}_1$$\rightarrow$0$^{+}_2)$=
8.4(26)~e$^2$fm$^4$ and a monopole strength
$\rho^2$(E0~:~0$^{+}_2$$\rightarrow$0$^{+}_1)$
=~8.7(7)$\times$10$^{-3}$. Comparisons to shell model
calculations point towards prolate-spherical shape coexistence and a phenomenological two level mixing model is used to extract
a weak mixing between the two configurations.

\end{abstract}

\maketitle

\indent

'Magic' nuclei exhibit large gaps between the occupied and valence
orbits. They are cornerstones of the nuclear structure, being used
(i) to test our understanding of the nuclear forces which form these
gaps and (ii) to model more complicated systems having many valence
nucleons. While nuclei having $8$ and $20$ protons (or neutrons) can
be reproduced by modeling the atomic nucleus with an harmonic
oscillator potential, a spin-orbit interaction must be added to
describe heavier magic nuclei. This spin-orbit interaction strongly
binds nucleons having their angular momenta $\ell$ aligned with
their intrinsic spin value $s$, denoted as $\ell _\uparrow$. This
leads throughout the chart of nuclei to regular sequence of orbits
$\ell _\uparrow$, $(\ell-2) _\uparrow$, $(\ell-2) _\downarrow$,
$\ell _\downarrow$, with the so-called large spin-orbit gaps $14$,
$28$, $50$, $82$ and $126$ between the lowered $\ell _\uparrow$
orbit ($\ell$=2, 3, 4, 5 and 6) and the others. Generally, in
particular at the stability, these gaps are large enough to prevent
excitations between occupied and valence orbits and these magic
nuclei are spherical. However, as the orbits forming the gap are
separated by two units of angular momentum, quadrupole excitations
are likely to develop if for some reason the shell gap is reduced.
In this hypothesis, the development of quadrupole excitations
jeopardizes the rigidity of the spherical gap and conduct the
nucleus to deform. Consequently the doubly magic nuclei which have
proton and neutron spin-orbit shell closures could become vulnerable
to quadrupole excitations, as both protons and neutrons could act
coherently to deform the nucleus. So far \emph{the} prototypical
deformed nucleus composed of such a double spin-orbit shell-closure
is $_{14}^{42}$Si$_{28}$~\cite{Bast07}. At N=28 a gradual
development of deformation occurs between the spherical doubly magic
$_{20}^{48}$Ca$_{28}$ and the deformed $_{14}^{42}$Si$_{28}$.
In between these two extremes, i.e. in $_{16}^{44}$S$_{28}$,
competition between spherical and deformed shapes is expected to be
present leading to shape coexistence~\cite{mf,per00,Caur04}.
Depending on the strength of the quadrupole correlations induced by
the cross shell excitations either the spherical normal
configuration, or the deformed one, becomes the ground state while
the other configuration forms a low lying 0$^+_2$ state. Therefore
the discovery and characterization of this 0$^+_2$ state in
$^{44}$S$_{28}$ represent crucial information for understanding the
evolution of N=28 shell gap. The non spherical nature of the
$^{44}$S ground state was suggested by its short $\beta$ half-life and
weak neutron-delayed emission probability~\cite{Sorl93}, by the
low energy of the 2$^+_1$ state (1297(18)~keV), and the enhanced
reduced transition probability B(E2~:~2$^{+}_1$$\rightarrow$0$^{+}_1$)
of 63(18)~e$^2$fm$^4$~\cite{Glas97}. However the 2$^+_1$ and B(E2)
values are intermediate between a rigid rotor and a spherical
nucleus. It suggests a possible mixing of spherical and deformed
shapes which can be deduced by studying the properties of the $0_2^+$
isomer at 1365(1)~keV observed in~\cite{Grev04}. Already the
study of a $7/2^-$ isomer in $^{43}$S~\cite{Sar00,Gaud09} has
shed light on shape coexistence in the N$\approx$28 region.
Other cases of shape coexistence
around shell closures have been reported in~\cite{12Be,30Mg}.

The present
letter reports on the determination of the monopole strength
$\rho^2$(E0~:~0$^{+}_2$$\rightarrow$0$^{+}_1$) and the reduced
transition probability B(E2~:~2$^{+}_1$$\rightarrow$0$^{+}_2$) in
$^{44}$S, extracted from the measurement of the half-life and the
branching ratio between the E$0$ and E$2$ decay of the isomeric
$0^+_2$ state. These pieces of information were obtained by using
combined $\gamma$ and electron delayed-spectroscopy and are
used to demonstrate the shape coexistence in $^{44}$S.

\indent The experiment was carried out at the Grand Acc\'el\'erateur
National d'Ions Lourds \textsc{(ganil)} facility. A primary beam of
$^{48}$Ca at 60A$\cdot$MeV (I$\sim$2e$\mu$A) impinged onto a
138~mg/cm$^2$ Be target to produce neutron-rich fragments. They were
separated by the \textsc{lise3} spectrometer~\cite{LISE} using an
achromatic 100~mg/cm$^2$ Be degrader. The
magnetic rigidity was set to optimize the transmission of the
$^{44}$S nuclei, produced at a rate of 200~sec$^{-1}$, with a
momentum acceptance of $\pm$1.45\%. Fragments were identified on an
event by event basis by means of their energy loss and magnetic
rigidity (B$\rho$) values. The B$\rho$ was obtained from the position of
the fragments at the dispersive focal plane given by a multi-wire
proportional chamber (\textsc{caviar})~\cite{Cav09}. The selected
nuclei were implanted in a 125~$\mu$m kapton foil tilted at 20~degrees with respect to the
beam axis.
Before the  foil, a stack of Si detectors, including a
position-sensitive one, was used to
adjust the implantation depth and to reconstruct the position of the
ions in a plane perpendicular to the beam axis. A thick Si detector
located downstream to the implantation foil was used as veto. The
$\gamma$ and electron decay events were registered up to 20~$\mu$sec
after the implantation. Electrons were detected
in four cooled 45*45~mm$^2$, 4~mm thick Si(Li) detectors, placed 20~mm above and below the beam axis.
The $\gamma$-rays were measured by two clover Ge detectors of the
\textsc{exogam} array located on the side of the implantation foil,
at a distance of 25~mm to the beam axis.
The use of a parallel beam optics along 2 meters length enables to
derive the ion implantation profile on the kapton foil from the
position-sensitive Si detector.
This ion profile, the geometry of the detectors and that of the
chamber were used as ingredients in a \textsc{geant4} simulation to
derive the electron ($\epsilon_{e^-}$) and $\gamma$
($\epsilon_{\gamma}$) efficiencies. The simulated efficiencies
compare well with the ones obtained with calibrated sources of
$^{207}$Bi and $^{152}$Eu placed in calibration runs at 6 different
positions on the implantation foil. Using these comparisons,
$\epsilon_{\gamma}$=3.06(5)\% and $\epsilon_{e^-}$=13.3(6)\% were
adopted for a gamma-ray of 1329~keV and an electron of 1362.5~keV,
respectively~\cite{for09}.

\begin{figure}
\includegraphics[width=7.5cm]{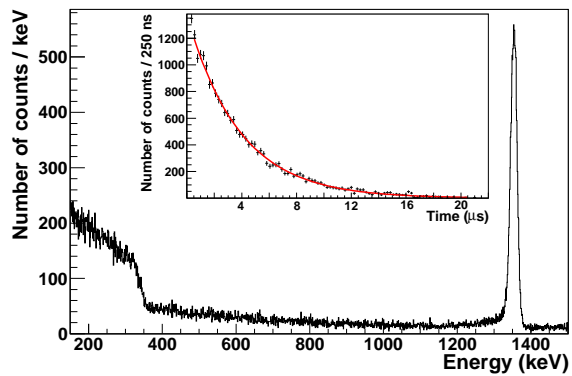}
\caption{\label{fig:fig_2} Electron energy spectrum obtained from
the Si(Li) detectors following the implantation of $^{44}$S nuclei.
The peak at 1362.5(1.0)~keV corresponds to the
0$^+_2$$\rightarrow$$0^+_1$ E0 transition. The low energy part is
due to pair creation. Insert: Time distribution of the 1362.5~keV
electron transition from which a half-life of 2.619(26)~$\mu$sec is
extracted.}
\end{figure}

The decay of the $0^+_2$ to the ground-state (E0) proceeds through
the emission of an internal conversion electron (IC) and by internal
pair formation (IPF). The electron spectrum, following the
implantation of a $^{44}$S nucleus, is shown in
Fig.~\ref{fig:fig_2}. A single peak is observed at 1362.5(10) keV
corresponding to an excitation energy of 1365(1) keV for the $0^+_2$
state, after having corrected for the binding energy of the K
electrons in the $^{44}$S nucleus. The integral of the peak is
$I_{e_{IC}^{-}}(E0)$= 148(8)$\times$10$^3$. The low energy part of
the spectrum is well accounted for by the pair formation (IPF)
mechanism in which electrons and positrons share an energy of
1365-(2*511)=343~keV.
The fit of the electron
time distribution
(insert of Fig.~\ref{fig:fig_2}) leads to an
half-life of 2.619(26)~$\mu$sec, which agrees with the value of
2.3(5)~$\mu$sec reported in~\cite{Grev04}.

\begin{figure}
\includegraphics[width=7.5cm]{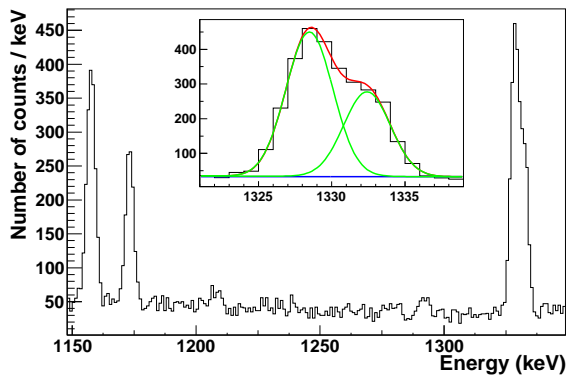}
\caption{\label{fig:fig_2b}
Part of the delayed gamma energy spectrum following the implantation of $^{44}$S nuclei.
Peaks from the $\beta$ decay of $^{44}$K (1158 keV) and $^{60}$Co (1173 and 1332.5~keV) are identified, the latter overlapping with the 1329~keV 2$^+_1$$\rightarrow$0$^+_1$ transition of $^{44}$S. The deconvolution of this doublet is shown in the insert.
}
\end{figure}

The 0$^+_2$$\rightarrow$2$^+_1$ decay branch, occurs through a
strongly converted E2 transition at 36(1)~keV, an energy below the
experimental threshold of the detection system.
The energy of this unobserved transition is derived from the
measured energies of the 0$^+_2$ and 2$^+_1$ states, the latter
being obtained from the observation of a delayed
2$^+_1$$\rightarrow$0$^+_1$ transition at 1329.0(5)~keV (half-life of 2.66(23)~$\mu$sec, 
in agreement with the value obtained from
the electron spectrum) which follows the
$0^+_2$$\rightarrow$$2^+_1$ decay. The 1329~keV energy agrees
with the value of 1297(18) found in Ref.~\cite{Glas97}. The yield of the
$0^+_2$$\rightarrow$$2^+_1$ transition, $I_{\gamma}(E2)$, has been
extracted from the number of delayed 2$^+_1$$\rightarrow$0$^+_1$
$\gamma$-rays.
As can be seen in the insert of
Fig.~\ref{fig:fig_2b}, this transition is contaminated by the 1332.5
keV $\gamma$-ray of $^{60}$Co arising from the activation of the 
last selection slits of the spectrometer, which also
produce a 1173 keV $\gamma$-ray with the same intensity. The number
of counts in the 1329 keV peak has been obtained by fitting the
$\gamma$ spectrum with two gaussians, the intensity of the 1332.5
keV transition being constrained by that of the 1173 keV
$\gamma$-ray. The resulting $I_{\gamma}(E2)$ is 56(3)$\times$10$^3$.

The decay of the $0^+_2$ state occurs through E2 and E0 transitions,
the ratio of which is expressed as :
\begin{equation}
R=\frac{\lambda(E2)}{\lambda(E0)}=
\frac{I_{\gamma}(E2)}{I_{e_{IC}^{-}}(E0)}
\frac{1+\alpha_{conv}(2^+_1\rightarrow0^+_1)}{1+\frac{\Omega_{IPF}}{\Omega_{IC}}}
\end{equation}
In this expression, the electronic factors for pair formation and
internal conversion have been extrapolated for a nucleus with A=44
from Ref.~\cite{a,b,c} to be
$\Omega_{IPF}$=1.495$\times10^7sec^{-1}$ and
$\Omega_{IC}$=1.1125$\times10^7sec^{-1}$, respectively. A value of
3.6$\times$10$^{-5}$ has been taken for the conversion coefficient
$\alpha_{conv}(2^+_1$$\rightarrow$0$^+_1)$ \cite{ban76}. Using the experimental
values of electron $I_{e_{IC}^{-}}(E0)$ and $\gamma$-ray
$I_{\gamma}(E2)$ yields derived above, the resulting branching ratio
is R=0.163(13). The $\rho^2$(E0~:~0$^+_2$$\rightarrow$$0^+_1$) and
B(E2~:~2$^+_1$$\rightarrow$0$^+_2$) values are obtained using the
following equations :
\begin{eqnarray}
\rho^2(E0)=\frac{ln(2)}{T_{1/2}(0^+_2)(1+R)(\Omega_{IPF}+\Omega_{IC})}\\
B(E2)=\frac{5.65\times10^{-10}}{5E_{\gamma}^5T_{1/2}(1+\frac{1}{R})
(1+\alpha_{conv}(0^+_2\rightarrow2^+_1)}
\end{eqnarray}
Using the measured branching ratio R, the half-life value
T$_{1/2}(0^+_2)$ and
$\alpha_{conv}(0^+_2$$\rightarrow$2$^+_1)$=10.94(1) extrapolated from Ref.~\cite{ban76}, the monopole strength
$\rho^2$(E0~:~0$^+_2$$\rightarrow$$0^+_1$) and the reduced
transition probability B(E2~:~2$^+_1$$\rightarrow$0$^+_2$) have been
determined
to be 8.7(7)$\times$10$^{-3}$ and 8.4(26)~e$^2$fm$^4$, respectively.

\vspace{0.1cm}
The values of  E(0$^+_2$)=1365(1)~keV and $\rho^2$(E0)=8.7(7)$\times$10$^{-3}$
are the smallest measured in this mass region, pointing to a weak
mixing between the 0$^+_1$ ground state and the 0$^+_2$ isomer and
therefore to shape coexistence. In case of a large mixing, these
states would repel each other to exhibit a large energy spacing and
a larger $\rho^2$(E0) value. To obtain further understanding on the
nature of the shape coexistence, data are compared to shell model
calculations.

\vspace{0.1cm}
Shell model (SM) calculations have been performed for
$^{44}_{16}$S$_{28}$ using the \textsc{antoine} code~\cite{ANTOINE1}
and the up-to-date \textsc{sdpf-u} interaction that accounts
remarkably well for nuclear structure in this mass
region~\cite{Nowa09}. The full $sd$ ($fp$) valence space has been
considered for protons (neutrons) using standard effective charges
$e_\pi$~=~1.35~$e$ ($e_\nu$~=~0.35~$e$).
\begin{table}
\caption{Experimental and shell model values for the excitation
energies, in MeV, and reduced transition probabilities B(E2), in
e$^2$fm$^4$, of $^{44}$S.}\label{tab:tab_1}
\begin{ruledtabular}
\begin{tabular}{lccc|cc}\noalign{\smallskip}
E/B(E2)& 2$^+_1$ & 0$^+_2$ & 2$^+_2$ & 2$^+_1$$\rightarrow$0$^+_1$ & 2$^+_1$$\rightarrow$0$^+_2$\\
\noalign{\smallskip}\hline\noalign{\smallskip}
exp. & 1.329(1) & 1.365(1) & 2.335(39) & 63(18) & 8.4(26)\\
SM & 1.172 & 1.137 & 2.140 & 75 & 19 \\
\noalign{\smallskip}
\end{tabular}
\end{ruledtabular}
\end{table}
\begin{figure} [h]
\includegraphics[height=6cm]{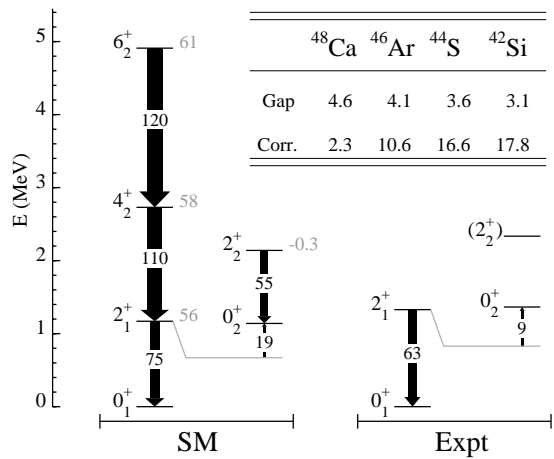}
\caption{\label{fig:fig_4} $^{44}$S level scheme calculated within the present SM approach (left), compared with available experimental data (right). E$2$ transition probabilities (in e$^2$fm$^4$) are reported on top of black arrows and intrinsic quadrupole moments (in $e fm^2$) are shown in light gray on the right side of calculated levels.
The ground state of the nucleus is head of a rotational band
($\beta\simeq$~0.25) and coexists
with the rather spherical low-lying $0^+_2$ isomer.
Calculated values of the N=28 gap and correlation energies (in MeV) are given for even-even N=28 isotones.
}
\end{figure}
The results gathered in Table~\ref{tab:tab_1} show a good agreement
with the experimental values, the only exception is a somewhat
larger calculated B(E2~:~2$^+_1$$\rightarrow$0$^+_2$) value than measured.
Nevertheless, both experiment and calculation agree with the fact
that the 2$^+_1$ state connects much strongly with the 0$^+_1$ state
than with the 0$^+_2$ one.
Indeed, the experimental B(E2~:~2$^+_1$$\rightarrow$0$^+_1$)/B(E2~:~2$^+_1$$\rightarrow$0$^+_2$) ratio is 7.5 whereas the calculated one is 3.2.
Calculated excited states connected to
these two 0$^+$ states are presented in Fig.~\ref{fig:fig_4} with
their intrinsic quadrupole moments $Q_0$. For sake of clarity only
the states of present interest are shown in this picture. Remarkable
is the presence of 2$^+_1$, 4$^+_2$ and 6$^+_2$ states on top of the
0$^+_1$ ground state connected by large B(E2) values. These states
present equal $Q_0$ values of about 60 $e fm^2$. These two features
characterize the presence of a rotational band from an axially
deformed nucleus with $\beta \simeq 0.25$. The 2$^+_{2}$ state at
2.14~MeV has a smaller intrinsic quadrupole moment $Q_0$=-0.3 $e
fm^2$ compatible with a spherical shape. A candidate for the 2$^+_2$
state is proposed at 2335(39)~keV by placing the previously reported 988~keV
transition~\cite{Soh02} on top of the 0$^+_2$ or 2$^+_1$ state.
Hence SM calculations suggest a prolate-spherical shape
coexistence in $^{44}$S.

A detailed analysis of the components contributing to the total
energy of the 0$^+$ states has been performed in order to deepen our
understanding on the evolution of the collectivity from
$^{48}_{20}$Ca to $^{42}_{14}$Si. Within the SM framework, the total
Hamiltonian can be separated into its monopole (i.e. spherical
mean-field contribution) and multipole (i.e. correlations mainly of
pairing and quadrupole type) parts~\cite{duf96}. As can be seen from
the values reported in Fig.~\ref{fig:fig_4}, correlations
strongly increase from the doubly magic $^{48}$Ca ($\simeq$2~MeV)
down to the exotic deformed $^{42}$Si ($\simeq$18~MeV), while the
size of the N=28 shell gap gets slightly reduced~\cite{gau07}.
This increase of correlations is favored on one hand by neutron
quadrupole excitations across the N=28 gap between the $f_{7/2}$
and $p_{3/2}$ orbits~\cite{gau07}, and on the other hand, by the degeneracy of
proton $s_{1/2}$ and $d_{3/2}$ orbits and excitations from the
$d_{5/2}$ shell~\cite{Bast07,ril08,gau10,mar10}.
In both cases, quadrupole correlations are favored by the fact that occupied and valence
states are separated by two units of angular momentum.
Without considering multipole contributions to the $0^+_1$ and
$0^+_2$ states in $^{44}$S, both levels are found to be
quasi-degenerate in energy, and the ground state of $^{44}$S is
spherical. A gain of 1.5~MeV from the multipole energy brings the
deformed configuration at the minimum of binding energy, while the
spherical configuration corresponds to the excited state. Similar
multipole effects energetically favor the oblate 0$^+$ state in
$^{42}$Si which is predicted to coexist with a prolate 0$^+$
state~\cite{Nowa09} at 1293~keV.

The shell model calculation uses an Harmonic Oscillator basis for the
description of the atomic nucleus. From the definition of the E0
operator, the calculated E0 transition between states of
the same harmonic oscillator shells, as for protons in the $sd$
shells and neutron in the $fp$ shells, is strictly zero. Therefore,
in order to shed light on the amount of mixing between the
0$^+_{1,2}$ states and to deduce their shape before mixing, we use a
phenomenological two interacting levels model. We assume two
spherical (S) and deformed (D) states before mixing which interact
to produce 0$^+_{1}$ and 0$^+_{2}$ states defined as :
\begin{eqnarray}
\vert0_1\rangle=cos\theta\vert0_D\rangle+sin\theta\vert0_S\rangle\\
\vert0_2\rangle=-sin\theta\vert0_D\rangle+cos\theta\vert0_S\rangle
\end{eqnarray}
where $\theta$ is the mixing angle. The E2 transition
between the 2$^+_1$ and 0$^+_2$ (or 0$^+_1$) states
being mainly due to the
D-component of these 0$^+$ states,
it follows that
B(E2~:~2$^+_1$$\rightarrow$0$^+_2$)/B(E2~:~2$^+_1$$\rightarrow$0$^+_1$)$\sim$tan$^2(\theta)$ (eq. 2 of~\cite{mac89}).
A mixing amplitude tan$^2(\theta)$=0.13 is deduced from the experimental B(E2) values whereas the shell model gives a somewhat larger value of 0.24, both being smaller than the case of a maximum mixing (tan$^2(\theta)$=1). Therefore, the shape coexistence is found to be more pronounced experimentally than calculated by the SM.
The magnitude of the monopole matrix
element can be written as a function of the mixing amplitude and of
the difference of shapes, $\beta_S$ and $\beta_D$, between the two
configurations before mixing~\cite{Wood99},
$\rho^2(E0)$=$(3Ze/4\pi)^2sin^2\theta cos^2\theta
(\beta_D^2-\beta_S^2)^2$. Using the experimental mixing amplitude
value of tan$^2(\theta)$ in this equation, the experimental monopole
strength is reproduced only when deformations $\beta_D \simeq$ 0.274
and $\beta_S$=0 are assumed. The deformation parameter $\beta_D$ is
in close agreement with the values obtained after mixing from
Coulomb excitation experiment~\cite{Glas97}, $\beta=0.258(36)$, and
from the shell model calculations, $\beta$=0.25. Altogether these
values again point towards a deformed-spherical shape coexistence in
$^{44}$S.

\indent To summarize, electron and $\gamma$ delayed-spectroscopies
have been used to determine the monopole strength
$\rho^2$(E0~:~0$^+_2$$\rightarrow$0$^+_1$)=8.7(7)$\times$10$^{-3}$
and the reduced transition probability
B(E2~:~2$^+_1$$\rightarrow$0$^+_2$)=8.4(26)~e$^2$fm$^4$ in the
$^{44}_{16}$S$_{28}$ nucleus. Using these values, the earlier
measured B(E2~:~2$^+_1$$\rightarrow$0$^+_1$)~\cite{Glas97}, shell
model calculations and a two level mixing model, it is found that
$^{44}$S exhibit a shape coexistence between a prolate ground state
($\beta \simeq$~0.25) and a rather spherical 0$^+_2$ state. This
establishes how the onset of collectivity progressively develops
between the spherical $^{48}_{20}$Ca and the deformed $^{42}_{14}$Si
nuclei. This study completes uniquely the understanding of the
shell-breaking mechanism at the spin-orbit closed-shell N=28,
which is as well of importance for the evolution of other shell gaps
having the same origin.

\indent We are grateful to the GANIL staff and the LISE team for
support. We acknowledge P. Van Isaker for fruitful discussions. This
work was supported by CNCSIS-UEFISCSU, proj. numb. PNII-IDEI
933/2007, Academy of Sciences of Czech Rep. and  by the European Community through
OTKA K68801.

\end{document}